*<u>Massive Multi-Omics Microbiome Database (M$^3$DB): A Scalable Data Warehouse and Analytics Platform for Microbiome Datasets</u>*

Shaun W. Norris[1]  (norrissw@vcu.edu)
Steven P. Bradley[2]  (bradleysp@vcu.edu)
Hardik I. Parikh[2]  (parikhhi@vcu.edu)
Nihar U. Sheth[1]*  (nsheth@vcu.edu)
*Corresponding author
[1]Center for the Study of Biological Complexity, Virginia Commonwealth University, Richmond, VA

[2]Department of Microbiology and Immunology, Virginia Commonwealth University, Richmond, VA

**Abstract:**
Massive Multi-Omics Microbiome Database (**M$^3$DB)** is a data warehousing and analytics solution designed to handle diverse, complex, and unprecedented volumes of sequence and taxonomic classification data obtained in a typical microbiome project using NGS technologies. M$^3$DB is a platform developed on Apache Hadoop, Apache Hive and PostgreSQL technologies. It enables users to store, analyze and manage high volumes of data, and also provides them the ability to query it in a fast and efficient manner. The M$^3$DB framework includes command line tools to process and store microbiome data, along with an easy-to-use web-interface for uploading, querying, analyzing and visualizing the data and/or results.

Availability: The source-code of M$^3$DB is freely available for download at http://www.github.com/nisheth/M3DB.

Contact: nsheth@vcu.edu

Supplementary Information:

## 1. Introduction

Modern sequencing technologies and tools have pushed bioinformatics in to the age of big data. As such traditional storage methods and relational database management systems used as data warehouses are inadequate. Storing information in text files takes a large amount of space, provides no ability to search and query, and requires large amounts of ram/CPU time in order to analyze. Distributed computing and MapReduce computer systems have proven to be an effective method for data warehousing and analysis of big data in other fields such as information technology, business informatics, social media, and many other applications (Qiu et al, 2010).

**M**assive **M**ulti-omics **M**icrobiome database (**M$^3$DB**) is a platform solution for the efficient storage/management and analysis of terabytes of microbiome sequencing data. At its core M$^3$DB is a framework that utilizes a custom API to process, analyze, and store sequence information from NGS and its taxonomic classifications. M$^3$DB uses Apache Hadoop, a distributed file system (HDFS) specifically optimized for big-data, to store/manage sequence and classification data, while microbial community profiles and additional metadata are stored in PostgreSQL. Apache Hive and PostgreSQL interact with HDFS for simplified data summarization, query and analysis. M$^3$DB provides user-friendly command line tools to upload, process and store microbiome data, while a web interface will provide users with an easy-to-use interactive platform to visualize and analyze raw data and/or results.

## 2. M³DB Architecture

The M³DB framework is written in Python and is centered on Apache Hadoop, Apache Hive and PostgreSQL. It is composed of three integrated layers. First, M³DB command-line tools (built using custom-APIs) can be used to create/upload metadata (related to projects/experiments/samples), upload sequence data and run analysis. Second, the M³DB database layer stores metadata and microbial community profile information in a PostgreSQL relational database. The sequences and sequence-level taxonomic classification information is stored using Apache Hadoop, which provides a distributed file system (HDFS) for optimal storage/management of large volumes of data. The detailed database schema is shown in Supplementary Information (Figure S1). M³DB uses Apache Hive (Hive Query Language HiveQL) that supports querying data stored in a Hadoop's HDFS, and the Python library psycogpg2 that communicates with the PostgreSQL database. The third layer, M³DB web-interface, is built using Django that allows users to query and visualize sample– and sequence–specific QC metrics, microbial abundance profiles and taxonomic classifications.  Figure 1 illustrates the M³DB architecture and how various components interact with each other.

## 3. Discussion

Here we illustrate the typical M³DB workflow.

1) Create/Upload Project, Experiment, Sample nodes using the APIs or web-interface. Data gets stored in traditional PostgreSQL database.

2) Upload sequence-data (typically in FASTQ format) using the APIs or web-interface. Data gets stored in HDFS.

3) Run Analysis using APIs or web-interface.

M³DB provides users with an option to quality filter the raw sequences using either Q-score or error probabilities. For paired-end sequences, M³DB runs MeFiT (Parikh et al.,2015) to merge and quality filter the reads. The quality metrics and the high quality reads (merged, if paired-end) get stored in HDFS. At present, M³DB supports taxonomic classification using tools like RDP (Wang et al., 2007) and STIRRUPS (Fettweis et al., 2012). Sample-specific abundance profiles are stored in PostgreSQL, while sequence-specific taxonomic classifications are stored in HDFS. The schematics of PostgreSQL database tables and use of Hadoop makes M³DB architecture extremely flexible to support other popular microbiome analysis tools like QIIME (Caporaso et al., 2010), Mothur (Schloss et al., 2009), USEARCH (Edgar, 2010), or Kraken (Wood and Salzberg, 2014).

4) Query and explore sample-specific statistics using Django web-interface.

## 4. Conclusions

To our knowledge, M$^3$DB is the only microbiome data management and analysis platform that utilizes modern big data technologies like Apache Hadoop and Apache Hive for warehousing sequences and metagenomic datasets. Compared to traditional databases, M$^3$DB provides highly scalable and resilient data storage and query functionality. It provides the utility to efficiently query large datasets, as well as the ability to streamline the process of merging, filtering and taxonomic classification. Future releases of M$^3$DB will include the implementation of tools like QIIME, mothur, Kraken, USEARCH. The web-interface will receive more visualization tools.

**Acknowledgements**
SN, SB and HP are supported by the NIH Common Fund Human Microbiome Project (HMP) program through grant 8U54HD080784 to G Buck, J Strauss, and K Jefferson. We gratefully acknowledge the technical and philosophical discussions with our colleagues Myrna Serrano, Paul Brooks, Jen Fettweis, Gregory Buck, and other members of the Vaginal Microbiome Consortium (VMC) at VCU.

**Figures:**

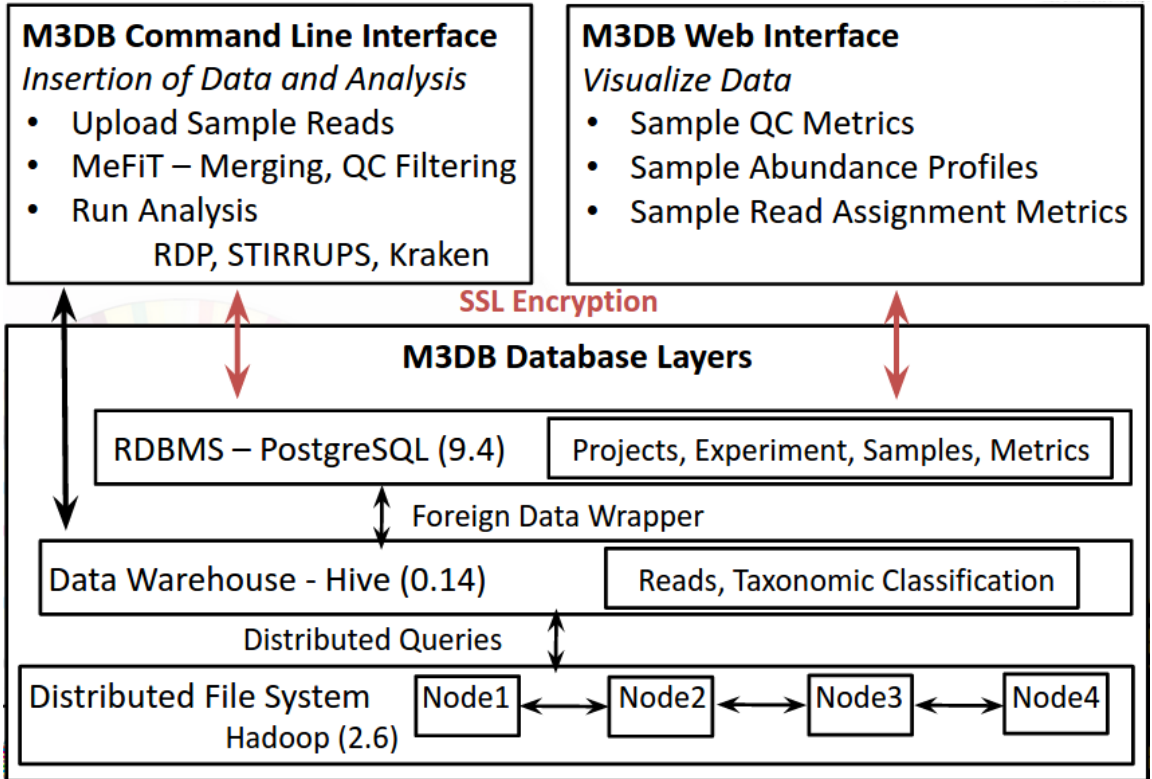

**Figure 1** – M³DB Architecture

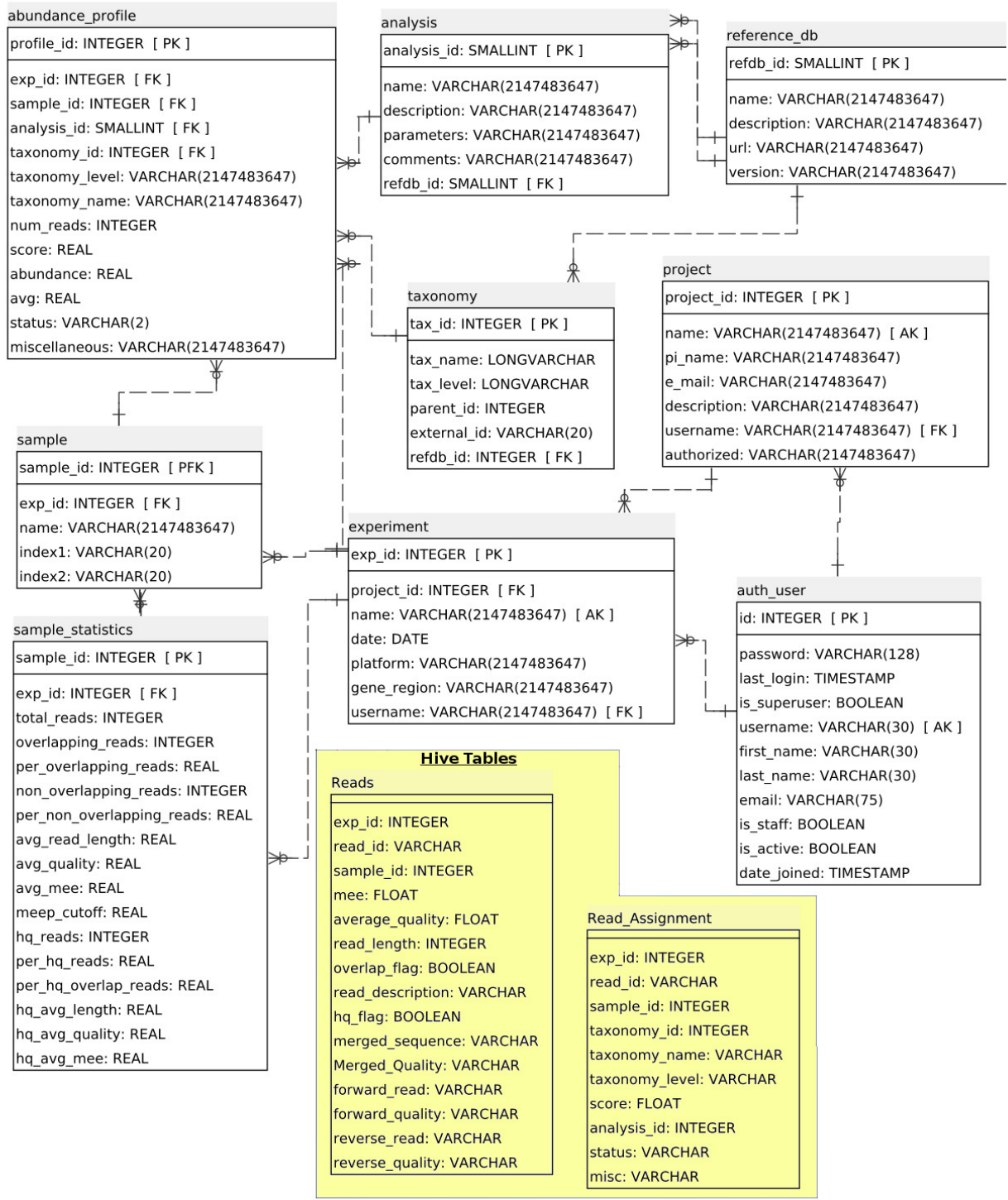

**Figure S1** - M³DB Database Schema